\documentclass[aps,showkeys,column,showpacs,preprintnumbers,amsmath,amssymb,superscriptaddress,floatfix,nofootinbib]{revtex4}
\usepackage{xcolor}
\usepackage{graphicx,color,dcolumn,booktabs,bm}
\usepackage{longtable,lscape}
\usepackage{txfonts}
\usepackage{overpic}
\usepackage{epsfig}
\usepackage{amssymb}
\usepackage{rotating}
\usepackage{epstopdf}
\usepackage{indentfirst}
\usepackage{feynmf}   
\usepackage{slashed}  
\usepackage{cases}
\usepackage{color}
\usepackage{multirow}
\usepackage{graphicx,color,dcolumn,booktabs,bm}
\usepackage{cases}
\usepackage{array}

\begin{document}


\title{Rotation and vibration in tetraquarks}

\author{Amir Jalili}
\email[]{jalili@fuw.edu.pl}
\affiliation{Institute of Theoretical Physics, Faculty of Physics, University of Warsaw, ul. Pasteura 5, PL-02-093 Warsaw, Poland}

\author{Jorge Segovia}
\email[]{jsegovia@upo.es}
\affiliation{Dpto. Sistemas F\'isicos, Qu\'imicos y Naturales, Univ. Pablo de Olavide, 41013 Sevilla, Spain}

\author{Feng Pan}
\affiliation{Department of Physics, Liaoning Normal University, Dalian 116029, P.R. China}
\affiliation{Department of Physics and Astronomy, Louisiana State University, Baton Rouge, LA 70803-4001, USA}

\author{Yan-An Luo}
\affiliation{School of Physics, Nankai University, Tianjin 300071, P.R. China}

\date{\today}

\begin{abstract}
A novel approach is introduced for obtaining precise solutions of the pairing Hamiltonian for tetraquarks, which utilizes an algebraic technique in infinite dimensions. The parameters involved in the transition phase are calibrated based on potential tetraquark candidates derived from phenomenology. Our investigation shows that the rotation and vibration transitional theory delivers a more accurate explanation for heavy tetraquarks compared to other methods utilizing the same formalism. To illustrate the concept, we compute the spectra of several tetraquarks, namely charm, bottom, bottom-charm and open charm and bottom systems, and contrast them with those of other particles.
\end{abstract}


\maketitle


\section{Introduction}
\label{sec:introduction}
Scientists have detected a new particle, dubbed X(2900), by analyzing all the data collected so far by the LHCb experiment at CERN's Large Hadron Collider \cite{lhcb1,lhcb2}. This experiment is renowned for discovering exotic quark combinations, which help scientists study the strong force, one of the four fundamental forces in the universe. LHCb has identified several tetraquarks, made up of four quarks (or two quarks and two antiquarks), including the latest discovery of an entirely new type of tetraquark with a mass of $2.9$ GeV/c$^2$, which has only one charm quark. While scientists predicted this particle's existence in 1964, it is the first observed instance of a tetraquark with only one charm quark. Quarks cannot exist independently; they form composite particles, such as mesons (a quark and an antiquark) or baryons (three quarks or three antiquarks), like the proton.
The LHCb detector located at the LHC focuses on studying $B$ mesons, which are composed of a bottom or an anti-bottom quark. These mesons quickly decay into lighter particles shortly after being produced in proton-proton collisions at the LHC. Tetraquarks are believed to be pairs of distinct mesons that are temporarily bound together like a "molecule," according to some theoretical models, while others view them as a single cohesive unit of four particles. Identifying and measuring the properties of new kinds of tetraquarks, such as their quantum spin and parity, will provide a better understanding of these strange inhabitants of the subatomic realm. The recently discovered particle, called X(2900), contains an anticharm, an up, a down, and an antistrange quark ($\bar{c}ud\bar{s}]$) and is considered the first open-charm tetraquark, as all previous tetraquark-like states observed by LHCb had a charm-anticharm pair, resulting in a net-zero "charm flavour." \cite{lhcb1,lhcb2}

Pairing interactions between fermionic or bosonic systems are common in many physical contexts such as Bose-Einstein Condensation and Superfluidity, airing correlations in nuclei: from microscopic to macroscopic models, high-temperature superconductors, ~\cite{pit,fra,leg, arim,oss91, oss93}. One example of the application of algebraic methods in hadron physics is the use of such interactions. \cite{bar,kru,b2,b1,f1, f2}. We establish explicit extensions of duality relations that relate the Hamiltonians and basis classification schemes associated with number-conserving unitary and number-nonconserving quasispin algebras for four-level pairing interactions. The Hamiltonian of the model can be defined using a linear combination of first- and second-order Casimir operators when one- and two-body interactions are present. The four-level pairing model describes a finite system that undergoes a second-order quantum phase transition between the rotation and vibration limits. Recently, we utilized the interacting boson approximation proposed by Arima and Iachello \cite{arima75,casten} to calculate wave functions in an interacting $sl$ many-body boson system~\cite{pan2002}. It is important to note that, in general, the building blocks of the boson system are associated with both $s$ and $l$ bosons for single and quadrupole angular momentum.
The bosonic pairing systems exhibit similarities in their Lie algebraic properties, but the differences are significant in terms of the irreducible representations (irreps) that the eigenstates transform under, which play a critical role in defining the system's spectroscopy. Finite pairing systems can be described by two complementary algebraic formulations: (1) a unitary algebra consisting of bilinear products of a creation and annihilation operator, and (2) a quasispin algebra that uses creation and annihilation operators for time-reversed pairs of particles \cite{hu,pan2006,cap8,27,me16,ajm17,epja}.

Tetraquarks are exotic hadrons composed of four quarks that can include two quarks and two antiquarks or four quarks of the same flavor. Despite being first proposed in the 1960s, their existence was only confirmed in 2013 by the Large Hadron Collider experiments. In recent years, the study of tetraquarks has gained increasing interest due to their unique properties and potential implications in particle physics ~\cite{f1}. In this context, we propose to apply an algebraic framework to investigate the properties of heavy tetraquarks $[QQ][\bar{Q}\bar{Q}]$. Our approach is based on the $SU(1,1)$ algebraic technique ~\cite{pan2006, ajm2021,ajmpt} and extends the $sl$ boson system. We will derive a new solvable model for hadron physics that takes into account the vector quark pairing strengths and examine the mass spectra of tetraquarks.

In recent years, there has been a growing interest in exploring the properties of fully-heavy tetraquarks. Theoretically, several models have been proposed to describe these states, including the diquark-antidiquark model, the chromomagnetic interaction model~\cite{Karliner:2016zzc}. On the experimental side, various searches have been performed to identify fully-heavy tetraquarks in high-energy experiments. For instance, the LHCb collaboration searched for deeply bound $bb\bar b \bar b$ tetraquark states, but no significant excess was found in the $\mu^+\mu^-\Upsilon(1S)$ invariant-mass distribution~\cite{Aaij:2018zrb}. However, the CMS experiment reported a potential candidate of a fully bottom tetraquark $T_{4b}=[bb][\bar{b}\bar{b}]$ around 18-19 GeV~\cite{dur}. Moreover, the LHCb collaboration has recently reported the observation of a narrow peak and a broad structure in the $J/\psi$-pair invariant mass spectrum, which could originate from hadron states consisting of four charm quarks~\cite{1804391}. These experimental results provide valuable information for further theoretical investigations of fully-heavy tetraquarks.

Various theoretical models have been developed to study fully-heavy tetraquarks, including phenomenological mass formulae~\cite{Karliner:2016zzc, Berezhnoy:2011xn, Wu:2016vtq}, QCD sum rules~\cite{Chen:2016jxd, Wang:2017jtz, Wang:2018poa, Reinders:1984sr}, QCD motivated bag models~\cite{Heller:1985cb}, NR effective field theories~\cite{Anwar:2017toa, Esposito:2018cwh}, potential models~\cite{Ader:1981db, Zouzou:1986qh, Lloyd:2003yc, Barnea:2006sd, Richard:2018yrm, Richard:2017vry, Vijande:2009kj, Debastiani:2017msn, Liu:2019zuc, Chen:2019dvd, Chen:2019vrj, Chen:2020lgj, Wang:2019rdo, Yang:2020rih}, non-perturbative functional methods~\cite{Bedolla:2019zwg}, and even some exploratory lattice-QCD calculations~\cite{Hughes:2017xie}. Some models predict that $QQ\bar Q\bar Q$ ($Q=c$ or $b$) bound states exist and have masses slightly below the respective thresholds of quarkonium pairs (see, for example, Refs.\cite{Chen:2016jxd, Anwar:2017toa, Karliner:2016zzc, Berezhnoy:2011xn, Wang:2017jtz, Wang:2018poa, Debastiani:2017msn, Esposito:2018cwh}). However, other studies suggest that no stable $cc\bar c\bar c$ and $bb\bar b\bar b$ tetraquark bound states exist because their masses are larger than two-quarkonium thresholds (see, for example, Refs.\cite{Ader:1981db, Lloyd:2003yc, Richard:2018yrm, Wu:2016vtq, Hughes:2017xie}). A better understanding of the mass locations of fully-heavy tetraquark states is crucial for our comprehension of their underlying dynamics and for experimental studies.


\section{Theoretical method}
\label{sec:theory}

In the context of describing a tetraquark system, diquark clusters play an important role. It is suggested that a tetraquark system, denoted by $T=Q_1Q_2\bar{Q}_3\bar{Q}_4$, consists of two point-like diquarks. To consider multi-level pairing in this context, we extend the interacting boson model using algebraic solutions of an $sl$-boson system~\cite{pan2002}. The dynamical symmetry group in this case is generated by $s$ and $l$ operators, where $l$ represents the configuration of the multiquark states. In the Vibron Model, scalar $s$-bosons with spin and parity $l^\pi=0^+$ and vector $l$-bosons with spin and parity $l^\pi=1^-$ represent elementary spatial excitations. The generators in the finite-dimensional $SU(1,1)$ algebra satisfy the following commutation relations.

\begin{subequations}
\begin{align}
\label{2a}
[S^{0}(l),S^{\pm}(l)] &= \pm S^{\pm}(l) \,, \\
\label{2b}
[S^{+}(l),S^{-}(l)] &= -2S^{0}(l).
\end{align}
\end{subequations}

We can use the $\widehat{SU(1,1)}$ algebra to describe the rotation and vibration transitional Hamiltonian of the $T_{4c}=[cc][\bar{c}\bar{c}]$, $T_{4b}=[bb][\bar{b}\bar{b}]$, and $T_{2bc}=[bc][\bar{b}\bar{c}]$ systems. It is worth mentioning that the quasi-spin algebras have been extensively discussed in previous studies, such as Refs.~\cite{pan2002, ajm17}.

Taking into account the generators of the $SU^{l}(1,1)$-algebra for tetraquarks given by Eqs.\eqref{2a} and\eqref{2b}, we can express the relevant quantities as linear combinations of these generators.
\begin{subequations}
\begin{align}
S^{+}({l}) &= \frac{1}{2} \, l^{\dag} \cdot l^{\dag} \,, \\
S^{-}({l}) &= \frac{1}{2} \, {\tilde{l}} \cdot {\tilde{l}} \,, \\
S^{0}({l}) &= \frac{1}{2} \, \left( l^{\dag} \cdot {\tilde{l}} + \frac{2l+1}{2} \right) \,,
\end{align}
\end{subequations}
where $l^{\dag}$ is the creation operator of an \emph{l}-boson constituting the tetraquark, and $\tilde{l}_\nu=(-1)^{\nu}l_{-\nu}$.

A complementary relation for tetraquark states can be expressed by
\begin{align}
|N; n_l\, \nu_l\,, n_\Delta JM\rangle =|N; \kappa_l\, \mu_l\,, n_\Delta JM\rangle \,,
\end{align}
with $\kappa_l=\frac{1}{2}\nu_l+\frac{1}{4}(2l+1)$ and $\mu_l=\frac{1}{2}n_l+\frac{1}{4}(2l+1)$, where $N$, $n_l$, $\nu_l$, $J$ and $M$ are quantum numbers of $U(N)$, $U(2l+1)$, $SO(2l+1)$, $SO(3)$ and $SO(2)$, respectively. The quantum number $n_\Delta$ is an additional one needed to distinguish different states with the same $J$.

The infinite dimensional $\widehat{SU(1,1)}$ Lie algebra is defined by
\begin{subequations}
\begin{align}
\label{4a}
S_{n}^{\pm} &= c_{Q_1}^{2n+1} S^{\pm}(l_1) + c_{Q_2}^{2n+1} S^{\pm}(l_2) + c_{\bar{Q}_3}^{2n+1} S^{\pm}(\bar{l}_3) \nonumber \\
&
+ c_{\bar{Q}_4}^{2n+1} S^{\pm}(\bar{l}_4), \\
\label{4b}
S_{n}^{0} &= c_{Q_1}^{2n} S^{0}(l_1) + c_{Q_2}^{2n} S^{0}(l_2) + c_{\bar{Q}_3}^{2n} S^{0}(\bar{l}_3) \nonumber \\
&
+ c_{\bar{Q}_4}^{2n} S^{0}(\bar{l}_4) \,,
\end{align}
\end{subequations}
The real-valued control parameters $c_Q$ and $c_{\bar{Q}}$ play a crucial role in determining the properties of tetraquarks. Specifically, $l_1$ and $l_2$ correspond to the first and second tetraquarks, respectively, while $\bar{l}_3$ and $\bar{l}_4$ correspond to the third and fourth tetraquarks. Additionally, the integer $n$ can take on values of $1$, $2$, $3$, and so on.

To ensure that the fully-heavy tetraquarks satisfy the correct properties, we impose the condition $S^{-}(l)|lw\rangle=0$ on the lowest weight state. The state $|lw\rangle$ can be defined as follows:
\begin{align}
|lw\rangle=|{N};\kappa_{l}\,\mu_{l}, n_\Delta JM \rangle,
\end{align}
where $ N=2k+ \nu_{Q_1}+ \nu_{Q_2}+ \nu_{\bar{Q}_3}+ \nu_{\bar{Q}_4}$.
 Hence, we have
\begin{equation}
S_{n}^{0} |lw\rangle =\Lambda_{n}^{l} |lw\rangle , \,\,\, \Lambda_{n}^{l}=\sum_l c_{l}^{2n}\frac{1}{2} \left(n_{l}+\frac{2l+1}{2} \right).
\end{equation}

The system shows vibrational and rotational transitions due to continuous variations of the pairing strengths, $c_l$, in the closed interval $[0,1]$. The all-heavy tetraquark pairing model undergoes a quantum phase transition. The vibration limit is reached when ${c_{Q_1} = c_{Q_2} = c_{\bar{Q}3} = c_{\bar{Q}4} = 0}$, while the rotational limit is attained when ${c_{Q_1} = c_{Q_2} = c_{\bar{Q}3} = c_{\bar{Q}4} = 1}$. In our analysis, we obtained diverse values for the control parameters, $c_{Q_i}$ and $c_{\bar{Q}_i}$, in the interval $[0,1]$ with $i=1,\ldots,4$, between the two limits.

The Hamiltonian of the heavy tetraquark pairing model is expressed in terms of the Casimir operators ${\hat C_2}$ using branching chains. The first two terms of the Hamiltonian, $S_{0}^{+} S_{0}^{-}$ and $S_{1}^{0}$, are associated with the $SU(1,1)$ algebra, while the remaining terms are constant in terms of the Casimir operators. In the duality relation for tetraquarks, the irreducible representations simplify the quasi-spin algebra chains~\eqref{4a} and~\eqref{4b}, and the labels for the chains are related via the duality relations. The Hamiltonian for the heavy tetraquark pairing model is derived by utilizing the generators of the $SU(1,1)$ algebra.
However, the pairing models of multi-level are also characterized by an overlaid $U(n_{1}+n_{2}+\ldots)$ algebraic structure with this branching:
${U(10)}_N \supset {SO(10)}_\nu \supset \mathop {SO(9)}_\nu   \supset \mathop {SO(3)}_s  \otimes \mathop {SO(3)}_{Q\,Q}  \otimes \mathop {SO(3)}_{\bar Q\,\bar Q}  \otimes \mathop {SO(3)}_J$

So, we can define the Hamiltonian with
\begin{align}
\label{a01}
\hat H &= g \, S_0^+ S_0^- + \alpha \, S_1^0 + \beta \, \hat{C}_2(SO(9)) \nonumber \\
&
+ \gamma_1 \, \hat{C}_2(SO(3)_R) + \gamma_2 \, \hat{C}_2(SO(3)_{Q_1Q_2}) \nonumber \\
&
+ \gamma_3\, \hat{C}_2(SO{(3)_{\bar Q_3\,\bar Q_4}}) + \gamma \, \hat{C}_2(SO{(3)_J}),
\end{align}
where $g$, $\alpha$, $\beta$, $\gamma _1$, $\gamma _2$, $\gamma _3$, and $\gamma$ are real-valued parameters.

To find the non-zero energy eigenstates with $k$-pairs, we exploit a Fourier Laurent expansion of the eigenstates of Hamiltonians which contain dependences on several quantities in terms of unknown $c$-number parameters $x_i$, and thus eigenvectors of the Hamiltonian for excitations can be written as
\begin{align}
&
|k;\nu_{Q_1}\nu_{Q_2}\nu_{\bar{Q}_3}\nu_{\bar{Q}_4} n_\Delta JM \rangle = \sum_{n_{i}\in Z} a_{n_{1}n_{2} \ldots n_{k}} \nonumber \\
&
=  x_{1}^{n_{1}} x_{2}^{n_{2}} x_{3}^{n_{3}} \ldots x_{k}^{n_{k}}S_{n_{1}}^{+} S_{n_{2}}^{+} S_{n_{3}}^{+} \ldots S_{n_{k}}^{+}|lw\rangle \,,
\label{wave}
\end{align}
and
\begin{align}
\label{quasi}
S _{n_{i}}^+ &= \frac {c_{Q_1}}{1-c_{Q_1}^{2} x_{i} } S^{+} (S_1)+\frac {c_{Q_2}}{1-c_{Q_2}^{2} x_{i} } S^{+} (S_2) \nonumber \\
&
+ \frac {c_{\bar{Q}_3}}{1-c_{\bar{Q}_3}^{2} x_{i} } S^{+} (\bar{S}_3)+\frac {c_{\bar{Q}_4}}{1-c_{\bar{Q}_4}^{2} x_{i} } S^{+}(\bar{S}_4) \,.
\end{align}
The coefficients $x_i$ are determined through the following set of equations
\begin{equation}
\label{ref-ref1}
\frac{\alpha}{x_{i}}=\sum_l \frac{ c_{l}^{2} (\nu_{l}+\frac{2l+1}{2})}{1-c_{l}^{2} x_{i}}-{\sum_{j\neq i}{\frac{2}{x_{i}-x_{j}}}} \,.
\end{equation}

In the pursuit of finding exact solutions for a spin-spin interaction system, Gaudin utilized a similar structure \cite{Gaudin76}  as an ansatz, which has now been verified as a consistent operator form in constructing the Bethe ansatz wavefunction for the present tetraquark system. To obtain the energy spectra, the Bethe ansatz equation (BAE), a non-linear equation, is employed for a $k$-pair excitation. The quantum number $k$-pair excitation pertains to the overall number of bosons $N$ and is linked to seniority numbers, specifically the quantum number $\nu_l$ of $SO(2l+1)$. As per equation (4), the allowed seniority numbers for a fixed $\nu_l$ include $n_l=\nu_l$, $\nu_l+2$, $\nu_l+4$, and so on. This information is well-established in the field.
Our approach to calculating the masses of heavy tetraquarks follows the procedure outlined in Ref. \cite{pan2006}. To account for the bosonic nature of the excitations (vibrations and rotations), we use the totally symmetric representation \eqref{a01} and define the boson number as the total number of vibrational states in the representation $[N]$.

Pairing in tetraquarks is an interesting phenomenon that affects their rotational and vibrational behavior. The quantum phase transition occurs between the vibrational and rotational limits in the fully-heavy tetraquark pairing model, and the quark (antiquark) configuration can undergo vibrations and rotations described by the quantum numbers $\nu_{Q_i}$, $\nu_{\bar{Q}_i}$, and $J$. While we will not consider bending and twisting in this analysis due to their higher mass requirements, we must account for the internal degrees of freedom of quarks and antiquarks.
To address this complication, we apply the method of pairing strengths, following Refs.\cite{f1, pan2002}. This scheme illustrates the stringlike configuration of the tetraquark and the vibration-rotation pattern we aim to identify. Within the two-quark configuration, we must follow the operator $Q$ with $\bar{Q}$, but since we are dealing with tetraquarks, we could also have combinations of $QQ$ and $\bar Q\bar Q$.
To determine the appropriate rotation-vibration pattern, we need to ensure that the pairing number $N = 2k + \nu_{Q_1} + \nu_{Q_2} + \nu_{\bar{Q}3} + \nu_{\bar{Q}_4}$ is satisfied. We can optimize the control parameters to find the exact symmetry of vibration and rotation that gives us the desired pairing number. By understanding the pairing behavior in tetraquarks, we can better understand their physical properties and potentially make predictions for future experiments.
In summary, our work builds on previous research in the field, but we introduce new ideas related to pairing in tetraquarks and optimize control parameters to identify the appropriate symmetry of vibration and rotation.


\section{Results}
\label{sec:results}

The determination of tetraquark mass in the diquark--anti-diquark pairing model requires solving the eigenvalue problem of Eq.~(\ref{a01}). However, in addition to the spins of diquark and antidiquark clusters, the $J^{PC}$ quantum numbers that define a tetraquark state also include the total spin, spatial inversion symmetry, and charge conjugation of the system.
Recent studies have shown that the $J^{PC}$ quantum numbers of a $Q_1Q_2\bar{Q}_3\bar{Q}_4$ system can be $0^{++}$, $1^{+-}$, and $2^{++}$, as discussed in Ref.~\cite{yang}. These quantum labels are essential for characterizing the properties of the tetraquark system.
The total spin of the tetraquark is determined by the combination of the spins of diquark and antidiquark clusters, and it affects the tetraquark's stability and decay properties. Spatial inversion symmetry is related to the tetraquark's mirror image, and it determines whether the system is symmetric or asymmetric with respect to spatial inversion. Charge conjugation, on the other hand, is related to the transformation of particles to their corresponding antiparticles and is a fundamental symmetry of the strong interaction.
Understanding the impact of total spin, spatial inversion symmetry, and charge conjugation on tetraquark states is crucial for predicting their properties and behavior. By considering these quantum numbers, we can gain insights into the tetraquark's internal structure and its interactions with other particles. This knowledge is essential for advancing our understanding of the strong interaction and the behavior of exotic hadrons. For scalar, vector and tensor systems, we have:

\begin{enumerate}
\item Two states for the scalar system:
\begin{subequations}
\begin{align}
|0^{++}\rangle &= |0_{Q_1Q_2}, 0_{\bar Q_3 \bar Q_4};J=0\rangle \,, \\
|0^{++\prime}\rangle &= |1_{Q_1Q_2}, 1_{\bar Q_3 \bar Q_4};J=0\rangle \,.
\label{eq:zeropp}
\end{align}
\end{subequations}
\item Three states for the vector system:
\begin{subequations}
\begin{align}
|A\rangle &= |0_{Q_1Q_2}, 1_{\bar Q_3 \bar Q_4};J=1\rangle \,, \\
|B\rangle &= |1_{Q_1Q_2}, 0_{\bar Q_3\bar Q_4};J=1\rangle \,, \\
|C\rangle &= |1_{Q_1Q_2}, 1_{\bar Q_3\bar Q_4};J=1\rangle \,.
\label{eq:onep}
\end{align}
\end{subequations}
Charge conjugation is a fundamental symmetry of the strong interaction that transforms particles into their corresponding antiparticles. This symmetry leads to different configurations in which $|A\rangle$ and $|B\rangle$ can interchange, while $|C\rangle$ remains odd.

In the $J^P=1^+$ configuration, we have one $C$-even and two $C$-odd states. This arrangement plays a crucial role in determining the properties and behavior of the system. Understanding the implications of these configurations is essential for predicting the tetraquark's stability and decay properties.
\begin{subequations}
\begin{align}
|1^{++}\rangle &= \frac{1}{\sqrt{2}}(|A\rangle+|B\rangle) \,, \\
|1^{+-}\rangle &= \frac{1}{\sqrt{2}}(|A\rangle-|B\rangle) \,, \\
|1^{+-\prime}\rangle &= |C\rangle \,.
\label{eq:onepp}
\end{align}
\end{subequations}
When considering tetraquarks, it is essential to choose appropriate values for the spin of the quark-antiquark pairs. In particular, the selection of spin states can impact the overall properties and behavior of the system.

In the case of a tetraquark composed of $Q_1, Q_2, \bar{Q}_3,$ and $\bar{Q}_4$, the appropriate spin states depend on the charge conjugation of the system.
Specifically, when $C=+$, the only allowed state is one where $Q_1\bar{Q}3$ has a spin of $S{Q_1\bar{Q}_3}=1$.
\item One state for the tensor system:
\begin{equation}
|2^{++}\rangle = |1_{Q_1Q_2}, 1_{\bar Q_3 \bar Q_4}; J=2\rangle \,,
\label{eq:twopp}
\end{equation}
where this state has also $S_{Q_1\bar Q_3}=1$.
\end{enumerate}


\subsection{The charm system}

The pairing tetraquark model considers two phases, rigid and non-rigid, which correspond to rotation and vibration symmetries, respectively. While both phases are idealized situations, they must coexist in reality, resulting in the emergence of vibrational-rotational modes in the transitional region. The parameters in this region are known as the phase parameters, where $c_{Q_i}=1$ with $i=1,\ldots,4$ corresponds to the rotational mode and $c_{Q_i}=0$ corresponds to the vibrational mode. The mass spectrum of the pairing tetraquark model can be calculated with fixed phase parameters, and the transitional spectra from one phase to another can be obtained by adjusting the phase parameters within the closed interval $[0,1]$.

To determine the phase coefficients, we can look at the meson-meson thresholds, such as $\eta_{c}(1S)\eta_{c}(1S)$ and $J/\psi(1S)J/\psi(1S)$ for $J^{PC}=0^{++}$, $\eta_{c}(1S)J/\psi(1S)$ for $J^{PC}=1^{+-}$, and $J/\psi(1S)J/\psi(1S)$ for $J^{PC}=2^{++}$, from a transitional theory perspective.

\begin{figure}[!t]
\includegraphics[width=0.475\textwidth]{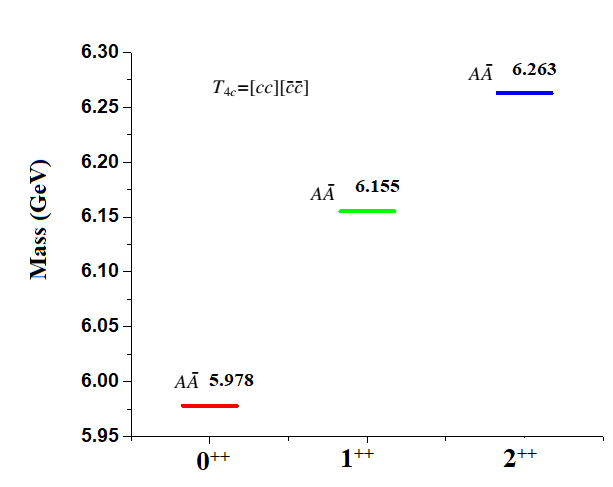}
\caption{\label{fig1} The predicted mass spectrum of the $T_{4c}$ tetraquarks. All spectroscopies are in GeV.}
\end{figure}

Our numerical values for the coefficients are $c_{Q_1}=0.92$, $c_{Q_2}=1$, and $c_{\bar{Q}_3}=c_{\bar{Q}_4}=0$. These values yield the following mass values:
\begin{align}
|0^{++\prime}\rangle &= |1_{cc}, 1_{\bar c \bar c};J=0\rangle: M=5.978\,\text{GeV} \,, \\
|1^{+-\prime}\rangle &= |1_{cc}, 1_{\bar c\bar c};J=1\rangle: M=6.155\,\text{GeV} \,, \\
|2^{++}\rangle &= |1_{cc}, 1_{\bar c \bar c}; J=2\rangle: M=6.263\,\text{GeV} \,,
\end{align}
for the $T_{4c}$ tetraquark system.
As shown in the (Fig.~\ref{fig1}) , we overall calculate the mass for  $T_{4c}$ tetraquark system.

\subsection{The bottom system}

The scenario presented here bears some resemblance to the earlier case. However, this time, according to the transitional theory, the extraction phase coefficients must be computed with regard to the meson-meson thresholds, such as $\eta_{b}(1S)\eta_{b}(1S)$ and $\Upsilon(1S)\Upsilon(1S)$ for $J^{PC}=0^{++}$, $\eta_{b}(1S)\Upsilon(1S)$ for $J^{PC}=1^{+-}$, and $\Upsilon(1S)\Upsilon(1S)$ for $J^{PC}=2^{++}$. Our numerical values for the coefficients are $c_{Q_1}=0.97$, $c_{Q_2}=1$, $c_{\bar{Q}3}=1$, and $c{\bar{Q}_4}=0$. These values yield the following mass values:
\begin{align}
|0^{++\prime}\rangle &= |1_{bb}, 1_{\bar b \bar b};J=0\rangle: M=18.752\,\text{GeV} \,, \\
|1^{+-\prime}\rangle &= |1_{bb}, 1_{\bar b\bar b};J=1\rangle: M=18.805\,\text{GeV} \,, \\
|2^{++}\rangle &= |1_{bb}, 1_{\bar b \bar b}; J=2\rangle: M=18.920\,\text{GeV} \,,
\end{align}
for the $T_{4b}$ tetraquark system.

\begin{figure}[!t]
\includegraphics[width=0.475\textwidth]{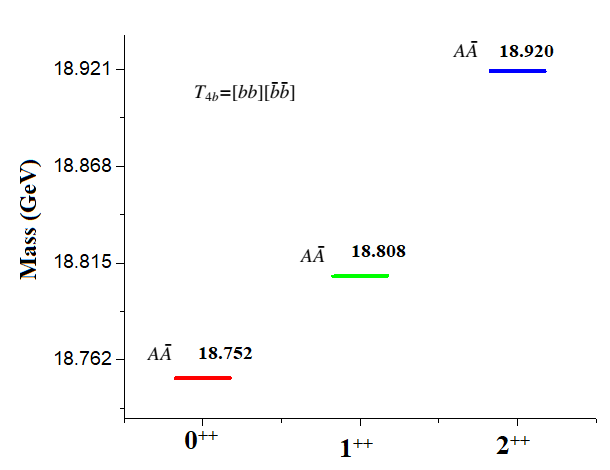}
\caption{\label{fig2} The predicted mass spectrum of the $T_{4b}$ tetraquarks. All spectroscopies are in GeV.}
\end{figure}

As shown in the (Fig.~\ref{fig2}) , we overall calculate the mass for  $T_{4b}$ tetraquark system.

\subsection{The bottom-charm system}

This study also considers the $T_{2bc}=[bc][\bar{b}\bar{c}]$ tetraquark structure, which combines $c$ quarks with $b$ quarks. Here, the $[bc]$ diquark spin may be either $0$ or $1$, allowing for the possibility of all states analyzed in the previous section. Once again, the best method for extracting the control parameters in $T_{2bc}$ tetraquarks, based on the transitional theory, is to utilize the corresponding meson-meson families. This method yields the following values: $c_{Q_1}=c_{Q_2}=1$, and $c_{\bar{Q}3}=c{\bar{Q}_4}=0$. The computed masses can be classified into the following categories:
\begin{itemize}
\item[(i)] The $J^{PC}=0^{++}$ contains two scalar states with masses
\begin{align}
|0^{++}\rangle &= |0_{bc}, 0_{\bar b \bar c};J=0\rangle: M=12.359\,\text{GeV} \,, \\
|0^{++\prime}\rangle &= |1_{bc}, 1_{\bar b \bar c};J=0\rangle: M=12.503\,\text{GeV} \,.
\end{align}
\item[(ii)] The $J^{PC}=1^{+-}$ contains two states with masses
\begin{align}
|1^{+-}\rangle &= \frac{1}{\sqrt{2}}(|0_{bc}, 1_{\bar b \bar c};J=1\rangle & \nonumber \\
&
-|1_{bc}, 0_{\bar b\bar c};J=1\rangle): M=12.896\,\text{GeV} \,, \\[1ex]
|1^{+-\prime}\rangle &= |1_{bc}, 1_{\bar b\bar c};J=1\rangle: M=12.016\,\text{GeV} \,.
\end{align}
\item[(iii)] The $J^{PC}=1^{++}$ contains one state with mass
\begin{align}
|1^{++}\rangle &= \frac{1}{\sqrt{2}}(|0_{bc}, 1_{\bar b \bar c};J=1\rangle \nonumber \\
&
+ |1_{bc}, 0_{\bar b\bar c};J=1\rangle): M=12.155\,\text{GeV} \,.
\end{align}
\item[(iv)] The $J^{PC}=2^{++}$ contains one state with mass
\begin{align}
|2^{++}\rangle = |1_{bc}, 1_{\bar b \bar c}; J=2\rangle: M=12.897\,\text{GeV} \,.
\end{align}
\end{itemize}

\begin{figure}[!t]
\includegraphics[width=0.475\textwidth]{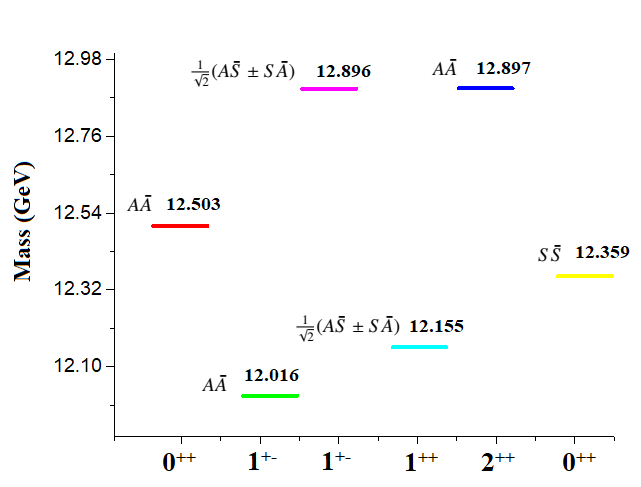}
\caption{\label{fig3} The predicted mass spectrum of the $T_{2bc}$ tetraquarks. All spectroscopies are in GeV.}
\end{figure}

As shown in the figure (Fig.~\ref{fig3}) , we overall calculate the mass for  $T_{2bc}$ tetraquark system.

\subsection{The open charm and bottom system}
In the most recent research, a thorough investigation was conducted on open charm (OC) and bottom (OB) tetraquarks comprising bottom and charm quarks, including $cq\bar{q}\bar{q}$, $cq\bar{s}\bar{q}$, $cs\bar{s}\bar{q}$, and $cs\bar{s}\bar{s}$ for charm, and $bq\bar{q}\bar{q}$, $bq\bar{s}\bar{q}$, $bs\bar{s}\bar{q}$, and $bs\bar{s}\bar{s}$ for bottom. The outcomes for the masses of these tetraquraks are summarized in Figs. 1, and the available experimental data is compared in Table. 1. Based on the method used in this study, the resulting values are demonstrated in the Caption of Fig. 1.
Our numerical values are $c_{Q_1}=0.83$, $c_{Q_2}=1$, $c_{\bar{Q}_3}=0$ and $c_{\bar{Q}_4}=0$,  $c_{Q_1}=0.86$, $c_{Q_2}=1$, $c_{\bar{Q}_3}=0$ and $c_{\bar{Q}_4}=0$
$c_{Q_1}=0.89$, $c_{Q_2}=1$, $c_{\bar{Q}_3}=0$ and $c_{\bar{Q}_4}=0$ and $c_{Q_1}=0.92$, $c_{Q_2}=1$, $c_{\bar{Q}_3}=0$ and $c_{\bar{Q}_4}=0$ for open charms $cq\bar{q}\bar{q}$, $cq\bar{s}\bar{q}$, $cs\bar{s}\bar{q}$, and $cs\bar{s}\bar{s}$, respectively.
In contrast, numerical values are $c_{Q_1}=0.91$, $c_{Q_2}=1$, $c_{\bar{Q}_3}=0.15$ and $c_{\bar{Q}_4}=0$,  $c_{Q_1}=0.93$, $c_{Q_2}=1$, $c_{\bar{Q}_3}=0.15$ and $c_{\bar{Q}_4}=0$ $c_{Q_1}=0.93$, $c_{Q_2}=1$, $c_{\bar{Q}_3}=0.29$ and $c_{\bar{Q}_4}=0$ and $c_{Q_1}=0.93$, $c_{Q_2}=1$, $c_{\bar{Q}_3}=0.37$ and $c_{\bar{Q}_4}=0$ for open bottom $bq\bar{q}\bar{q}$, $bq\bar{s}\bar{q}$, $bs\bar{s}\bar{q}$, and $bs\bar{s}\bar{s}$, respectively.

\begin{figure*}[!t]
\includegraphics[width=1.0 \textwidth]{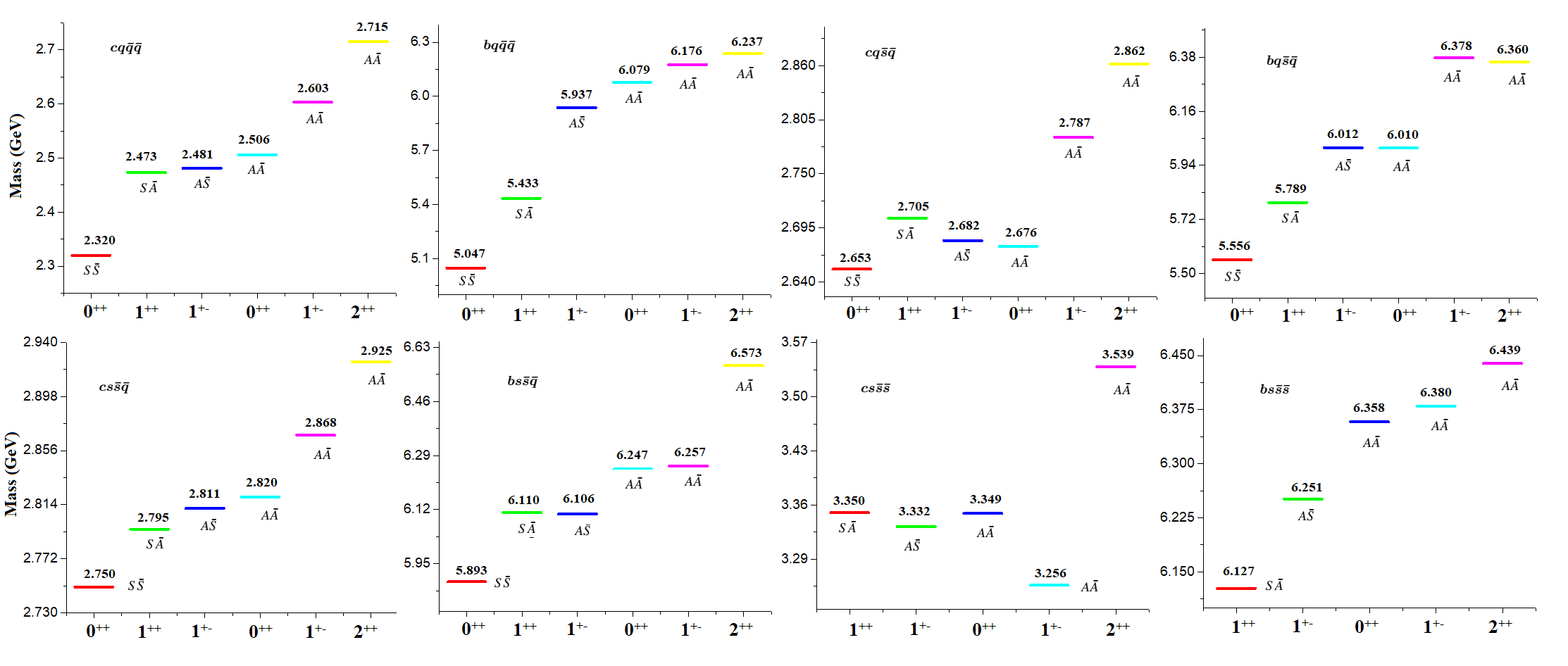}
\caption{\label{fig4} The predicted mass spectrum of the open charm and bottom tetraquarks. All spectroscopies are in GeV.}
\end{figure*}

As shown in the (Fig.~\ref{fig4}) , finally we calculate the mass for  open charm and bottom tetraquark systems.

\section{Discussion}
\label{sec:discussion}

\begin{table*}\label{t1}
\caption{\label{tab:QQmass} Masses of fully-heavy tetraquark systems as computed within the theoretical framework presented herein. The meson-meson threshold is $E_{\rm th}$, and $\Delta=M-E_{\rm th}$ represents the energy distance of the tetraquark with respected its lowest meson-pair threshold. The notation $s$ and $a$ indicates scalar and axial-vector diquarks.}
\begin{ruledtabular}
\begin{tabular}{ c c c c c c c }
Structure &Configuration   & $J^{PC}$ & $M_{\rm tetra}$ in this work (GeV) & Threshold & $E_{\rm th}$ (GeV)& $\Delta$ (GeV)
\\
\hline
\multirow{4}{*}{$T_{4c}$=$[cc] [\bar c \bar c]$} & \multirow{4}{*}{$A \bar A$} & \multirow{2}{*}{$0^{++}$} & \multirow{2}{*}{5.978} & $\eta_{c}(1S)\eta_{c}(1S)$ & 5.968 & 0.01
\\
\cline{5-7}
& & & &  $J/\psi(1S)J/\psi(1S)$ & 6.194 & -0.216
\\
\cline{3-7}
& & $1^{+-}$ & 6.155 & $\eta_{c}(1S)J/\psi(1S)$ & 6.081 & 0.074
\\
\cline{3-7}
& & $2^{++}$ & 6.263 & $J/\psi(1S)J/\psi(1S)$ & 6.194 & 0.069
\\
\cline{1-7}
\multirow{4}{*}{$T_{4b}$=$[bb][ \bar b \bar b]$} & \multirow{4}{*}{$A \bar A$} & \multirow{2}{*}{$0^{++}$} & \multirow{2}{*}{18.752} & $\eta_{b}(1S)\eta_{b}(1S)$ & 18.797 & -0.045
\\
\cline{5-7}
  & & & &  $\Upsilon(1S)\Upsilon(1S)$ & 18.920 & -0.168
\\
\cline{3-7}
  & & $1^{+-}$ & 18.808 & $\eta_{b}(1S)\Upsilon(1S)$ & 18.859 & -0.051
\\
\cline{3-7}
  & & $2^{++}$ & 18.920 & $\Upsilon(1S)\Upsilon(1S)$ & 18.920 & 0.0
  \\
  \cline{1-7}
  \multirow{21}{*}{$T_{2bc}$=$[bc][ \bar b \bar c]$} & \multirow{10}{*}{$A \bar A$} & \multirow{4}{*}{$0^{++}$} & \multirow{4}{*}{12.503} & $\eta_{b}(1S)\eta_{c}(1S)$ & 12.383 & 0.12
\\
\cline{5-7}
  & & & & $J/\psi(1S)\Upsilon(1S)$ & 12.557 & -0.054
\\
\cline{5-7}
  & & & & $B_{c}^{\pm}  B_{c}^{\mp}$ & 12.550 & -0.047
\\
\cline{5-7}
  & & & & $B_{c}^{*\pm} B_{c}^{*\mp}$ & 12.666 & -0.163
\\
\cline{3-7}
  & & \multirow{4}{*}{$1^{+-}$} & \multirow{4}{*}{12.016} & $\eta_{c}(1S)\Upsilon(1S)$ & 12.444 &-0.428
\\
\cline{5-7}
  & & & & $J/\psi(1S)\eta_{b}(1S)$ & 12.496 & -0.48
\\
\cline{5-7}
  & & & & $B_{c}^{\pm}  B_{c}^{*\mp}$ & 12.608 & -0.592
\\
\cline{5-7}
  & & & & $B_{c}^{* \pm}  B_{c}^{* \mp}$ & 12.666 & -0.65
\\
\cline{3-7}
  & & \multirow{2}{*}{$2^{++}$} & \multirow{2}{*}{12.897} & $J/\psi(1S)\Upsilon(1S)$ & 12.557 & 0.34
\\
\cline{5-7}
  & & & & $B_{c}^{* \pm}  B_{c}^{*\mp}$ & 12.666 & 0.231
\\
\cline{2-7}
  & \multirow{7}{*}{$\frac{1}{\sqrt{2}}(A \bar S \pm S \bar A)$} & \multirow{3}{*}{$1^{++}$} & \multirow{4}{*}{12.155} & $J/\psi(1S)\Upsilon(1S)$ & 12.557 & -0.402
\\
\cline{5-7}
  & & & & $B_{c}^{\pm} B_{c}^{*\mp}$ & 12.608 & -0.453
\\
\cline{5-7}
  & & & & $B_{c}^{*\pm}  B_{c}^{*\mp}$ & 12.666 & -0.511
\\
\cline{3-7}
  & & \multirow{4}{*}{$1^{+-}$} &  \multirow{4}{*}{12.896} & $\eta_{c}(1S)\Upsilon(1S)$ & 12.444 & 0.452
\\
\cline{5-7}
  & & & & $J/\psi(1S)\eta_{b}(1S)$ & 12.496 & 0.4
\\
\cline{5-7}
  & & & & $B_{c}^{\pm} B_{c}^{*\mp}$ & 12.608 & 0.288
\\
\cline{5-7}
  & & & & $B_{c}^{*\pm}  B_{c}^{*\mp}$ & 12.666 & 0.23
\\
\cline{2-7}
  & \multirow{4}{*}{$S \bar S$} & \multirow{4}{*}{$0^{++}$} & \multirow{4}{*}{12.359} & $\eta_{c}(1S)\eta_{b}(1S)$ & 12.383 & -0.024
\\
\cline{5-7}
  & & & & $J/\psi(1S)\Upsilon(1S)$ & 12.557 & -0.198
\\
\cline{5-7}
  & & & & $B_{c}^{\pm}  B_{c}^{\mp}$ & 12.550 & -0.191
\\
\cline{5-7}
  & & & & $B_{c}^{*\pm}  B_{c}^{*\mp}$ & 12.666 & -0.307
\\
\end{tabular}
\end{ruledtabular}
\end{table*}

\begin{table*}\label{t2}
\caption{\label{tab:cm1} Comparison of our results with theoretical predictions for the masses of $T_{4b}=[bb][\bar{b}\bar{b}]$, and $T_{4c}=[cc][\bar{c}\bar{c}]$ tetraquarks. All results are in GeV.}
\begin{ruledtabular}
\begin{tabular}{ccccccc}
Reference & \multicolumn{3}{c}{$bb \bar b \bar b$}& \multicolumn{3}{c}{$cc \bar c \bar c$}\\
\cline{2-4} \cline{5-7} & $0^{++}$ & $1^{+-}$ & $2^{++}$ & $0^{++}$ & $1^{+-}$ & $2^{++}$\\
\hline
 \centering{This paper} & 18.752 & 18.808 & 18.920& 5.978 & 6.155 & 6.263  \\
 \centering{\cite{p2}} & 18.460-18.490 & 18.320-18.540 & 18.320-18.530& 6.460-6.470  & 6.370-6.510 & 6.370-6.510  \\
  \centering{\cite{FullBeauty2019}}& 18.690  &- &-&-&-&- \\
  \centering{\cite{FullHeavy2019sec}} & 18.748 & 18.828 & 18.900& 5.883 & 6.120 & 6.246 \\
  \centering{\cite{FullHeavy2018}} & 18.750 &- &-&$ <6.140$ &- &-   \\
  \centering{\cite{D12},\cite{blln}}  & 18.754 & 18.808 & 18.916& 5.966 & 6.051 & 6.223  \\
 \centering{\cite{FullHeavy2017,Karliner:2020dta}} & $18.826 $ &- & $18.956$& $6.192$ & -&$6.429$ \\
 \centering{\cite{SumR2,E18}}& $18.840 $ & $18.840 $ & $18.850 $ & $5.990 $ & $6.050 $ & $6.090 $   \\
 \centering{\cite{Chen}}&19.178&19.226&19.236&-&-&-\\
 \centering{\cite{Jin:2020jfc}}&19.237&19.264&19.279&6.314&6.375&6.407\\
\centering{\cite{WLZ}}&19.247 &19.247  &19.249&6.425&6.425&6.432 \\
\centering{\cite{FullHeavy2019,liu:2020eha}} & 19.322 & 19.329 & 19.341& 6.487 & 6.500 & 6.524  \\
\centering{\cite{p5}}&19.329&19.373&19.387&6.407&6.463&6.486\\
 \centering{\cite{Lu:2020cns}}&19.255&19.251&19.262&6.542&6.515&6.543\\
  \centering{\cite{p1}} & 20.155 & 20.212 & 20.243& 6.797 & 6.899 & 6.956 \\
  \centering{\cite{FullCharm2017,E19}} &-&-&-& 5.969 & 6.021 & 6.115\\
  \centering{\cite{102}} &-&-&-& 6.695 & 6.528 & 6.573\\
  \centering{\cite{103}} &-&-&-& 6.480 & 6.508 & 6.565\\
  \centering{\cite{tetrababc}} & 19.666 & 19.673 & 19.680& 6.322 & 6.354 & 6.385 \\
   \centering{\cite{tetrac}} & - & - & -& 6.510 & 6.600 & 6.708 \\
   \centering{\cite{25}} & 18.981 & 18.969 &19.000& 6.271 & 6.231 & 6.287 \\
   \centering{\cite{26}} & 19.314 &19.320 &19.330& 6.190 & 6.271 &6.367 \\
   \centering{\cite{p4}}~set. I & 18.723 & 18.738 &20.243& 5.960 & 6.009 & 6.100 \\
   \centering{\cite{p4}}~set. II & 18.754 & 18.768 &18.797& 6.198 & 6.246 & 6.323 \\
   \centering{\cite{zha}} & 19.226 & 19.214 &19.232& 6.476 & 6.441 & 6.475 \\
  \end{tabular}
\end{ruledtabular}
\end{table*}

\begin{table*}\label{t3}
\caption{\label{tab:cm2} Comparison of our results with theoretical predictions for the masses of $T_{2bc}=[bc][\bar{b}\bar{c}]$  tetraquarks. All results are in GeV.}
\begin{ruledtabular}
\begin{tabular}{ccccccc}
\centering{Reference} & \multicolumn{3}{c}{{$A \bar A$}} & \multicolumn{2}{c}{{$\frac{1}{\sqrt{2}}(A \bar A \pm A \bar A)$}} & \multicolumn{1}{c}{{$S \bar S$}}
\\[1ex]
\cline{2-4} \cline{5-6} \cline{7-7}
 & \centering{$0^{++}$} & \centering{$1^{+-}$} & \centering{$2^{++}$} & \centering{$1^{++}$} & \centering{$1^{+-}$} & \multicolumn{1}{c}{$0^{++}$}
\\
\hline
\centering{This paper} & 12.503 & 12.016 & 12.897 & 12.155 & 12.896 & 12.359\\
  \centering{\cite{D12}} &12359 & 12424  & 12566 & 12485 &12488 & 12471\\
     \centering{\cite{FullHeavy2019sec}} & 12374 & 12491 & 12576 &12533 & 12533 & 12521\\
    \centering{\cite{FullHeavy2018}} & $<12620$ &- &- &- &- &-\\
   \centering{\cite{Chen2}}&12746&12804&12809&-&12776&-\\
   \centering{\cite{p5}}&12829&12881&12925&-&-&-\\
\centering{\cite{FullHeavy2019}} & 13035  & 13047 & 13070 & 13056 & 13052 & 13050\\
  \centering{\cite{p1}} & 13483 & 13520 & 13590 & 13510 & 13592 & 13553\\
\end{tabular}
\end{ruledtabular}
\end{table*}

\begin{table*}\label{to}
\caption{\label{tab:cm2} Comparison of our results with theoretical predictions for OC and OB tetraquark states with diquark-antidiquark in ground $1S$ state. All results are in GeV.}
\begin{ruledtabular}
\begin{tabular}{cccccccccc}
 $J^P$ &  Diquark content & Experiment \cite{pg} & Mass& OC \cite{lu} & OB \cite{lu} &  OC \cite{ebert} & OB  \cite{ebert} &OC (This work) &OB (This work) \\\hline
\\
&               &  Meson    &Mass & $\bm{cq\bar q \bar q}$      & $\bm{bq\bar q \bar q}$ & $\bm{cq\bar q \bar q}$      & $\bm{bq\bar q \bar q}$ &$\bm{cq\bar q \bar q}$      & $\bm{bq\bar q \bar q}$  \\
 $0^+$         &  $S\bar S$      &$D^*_0$(2.400) &$\frac{2.403}{2.318}$& 2.729                 & 6.063&  2.399   &5.758   &2.320  &5.047   \\
 $1^+$         &  $S\bar A$      & && 2.838                 & 6.077& 2.558    &5.950   &2.473  &5.433   \\
 $1^+$         &  $A\bar S$     &$D_1$(2.430) &2.427& 2.767                 & 6.164& 2.473    &5.782   &2.481  &5.937    \\
 $0^+$         &  $A\bar A$      & && 2.575                 & 6.046&2.503     &5.896   &2.506  &6.079   \\
 $1^+$         &  $A\bar A$      & && 2.747                 & 6.118& 2.580    &5.937   &2.603  &6.176   \\
 $2^+$         &  $A\bar A$      & && 2.969                 & 6.226&2.698     &6.007   &2.715  &6.237   \\\hline
               &            &&   & $\bm{cq\bar s \bar q}$      & $\bm{bq\bar s \bar q}$ & $\bm{cq\bar s \bar q}$      & $\bm{bq\bar s \bar q}$ &$\bm{cq\bar s \bar q}$      & $\bm{bq\bar s \bar q}$      \\
 $0^+$         &  $S\bar S$      &$D_s$(2.632) &2.6325& 2.873                 & 6.196&2.619     &5.997   &2.653  &5.556   \\
 $1^+$         &  $S\bar A$      & && 2.957                 & 6.210&2.723     &6.125   &2.705  &5.789   \\
 $1^+$         &  $A\bar S$      & && 2.911                 & 6.274&2.678     &6.021   &2.682  &6.012   \\
 $0^+$         &  $A\bar A$     & & & 2.692                 & 6.150&2.689     &6.086   &2.676  &6.010   \\
 $1^+$         &  $A\bar A$     & & & 2.866                 & 6.226&2.757     &6.118   &2.787  &6.378   \\
 $2^+$         &  $A\bar A$      &$D^*_{sj}$(2.860) &2.862& 3.087                 & 6.337&2.863     &6.177   &2.862  &6.360   \\\hline
               &               &      && $\bm{cs\bar s \bar q}$      & $\bm{bs\bar s \bar q}$ & $\bm{cs\bar s \bar q}$      & $\bm{bs\bar s \bar q}$& $\bm{cs\bar s \bar q}$      & $\bm{bs\bar s \bar q}$ \\
 $0^+$         &  $S\bar S$     & & & 3.001                 & 6.317&2.753     &6.108   &2.750  &5.893   \\
 $1^+$         &  $S\bar A$      & && 3.085                 & 6.330&2.870     &6.238   &2.795  &6.110   \\
 $1^+$         &  $A\bar S$      & && 3.035                 & 6.394 &2.830     &6.134   &2.811  &6.106   \\
 $0^+$         &  $A\bar A$     & && 2.827                 & 6.272&2.839     &6.197   &2.820  &6.247   \\
 $1^+$         &  $A\bar A$       & && 2.994                 & 6.347&2.901     &6.228   &2.868  &6.257  \\
 $2^+$         &  $A\bar A$      & && 3.207                 & 6.456&2.998     &6.284   &2.925  &6.573   \\\hline
               &              &     & & $\bm{cs\bar s \bar s}$      & $\bm{bs\bar s \bar s}$ & $\bm{cs\bar s \bar s}$      & $\bm{bs\bar s \bar s}$& $\bm{cs\bar s \bar s}$      & $\bm{bs\bar s \bar s}$  \\
 $1^+$         &  $S\bar A$      & && 3.201                 & $-$&3.025     &6.383   &3.350  &6.127   \\
 $1^+$         &  $A\bar S$      & && $-$                  & 6.504& -    &-  &3.332  &6.251   \\
 $0^+$         &  $A\bar A$     & & & 2.942                 & 6.376&3.003     &6.353    &3.349  &6.358   \\
 $1^+$         &  $A\bar A$      & && 3.111                 & 6.455&3.051     &6.372   &3.256  &6.380   \\
 $2^+$         &  $A\bar A$     & & & 3.322                 & 6.566&3.135     &6.411   &3.539  &6.439   \\
\end{tabular}
\end{ruledtabular}
\end{table*}

The calculation process involves a fixed set of Hamiltonian parameters while allowing the phase parameters to fluctuate during the transition. In Ref.\cite{f2}, the authors demonstrated that the boson number's quantum value could be obtained by taking the $N\to \infty$ limit. It was found that taking $N$ to be a large number was sufficient to account for all known and unknown states up to the maximum value of the quantum number of the angular momentum and other relevant quantum numbers for the applications. In this current research, we are utilizing the same approach as in Ref.\cite{f2}, setting $N=100$ to ensure that all states up to the maximum quantum number are considered.

The Hamiltonian's pattern is comparable to the $O(4)$ restriction proposed in mesons, where the control parameter is set to $1$. Our study indicates that the control parameters $c_{\bar{Q}3}$ and $c{\bar{Q}4}$ cannot be considered as 1 in the presence of heavy antiquarks, except for $T{4b}$ tetraquarks. This is due to the fact that the $T_{4b}$ tetraquark's mass is two to three times more substantial than that of $T_{2bc}$ and $T_{4c}$ tetraquarks. For heavy mass tetraquarks, pairing strength plays a significant role, as evidenced by the larger $c_{Q_1}$ value in the $T_{2bc}$ case than in the $T_{4c}$ case. The same reasoning applies to the open and bottom tetraquark system.

The Hamiltonian's parameters for the discussed structures are presented in the Figures' captions. During the transition phase, we set $\alpha$ to be 1.5. As the pairing model's vibrational-rotational transition is a second-order quantum phase transition, the masses of the wave functions in the tetraquark's vibrational model are smooth with respect to parameter changes. This enables us to determine them in the transition region.

In Table~\ref{tab:QQmass}, we present the difference between the calculated masses of the tetraquarks and the threshold for meson-pairing. The values of $\Delta$ represent the difference between the tetraquark mass $M_{\rm tetra}$ and its lowest meson-meson threshold $E_{\rm th}$. If $\Delta$ is negative, it implies that the tetraquark state lies below the fall-apart decay threshold and should therefore be stable. On the other hand, a state with a small positive $\Delta$ could be observed as a resonance due to suppression by phase space. The states with high positive $\Delta$ values are considered broad and difficult to detect in experimental analyses.

Our investigation indicates that slightly deviating the control parameter $c_{Q_1}$ from $1$ is more suitable for determining the tetraquark masses, especially for the extensive $T_{2bc}$ families. Additionally, for $T_{4b}=[bb][\bar{b}\bar{b}]$ states, the dominant contribution comes from the pairing of $c_{\bar{Q}3}$ and $c{\bar{Q}_4}$ quarks, indicating that phase parameters for $\bar{Q}_3$ and $\bar{Q}_4$ quarks become significant in computing tetraquark masses at high energy, around $18-19,\text{GeV}$. On the other hand, at low energy, there is a competition between ${Q}_1$ and ${Q}_2$.

The energy spectra of the fully-heavy tetraquarks under study, where $c_{Q_i}$ values are in the range of $0.9-1.0$, can be attributed to a rotational phase, based on the definition mentioned above.
It is worth noting that a change of $\pm15\%$ in all coefficients results in a maximum variation of $30\%$, $23\%$, and $17\%$ in the masses of the $T_{4c}$, $T_{4b}$, and $T_{2bc}$ tetraquark systems, respectively.
However, it should be emphasized that the masses of the remaining tetraquark states undergo lesser modifications.

Our study on the vibrational and rotational transitions in open and bottom tetraquarks suggests that a slight deviation of the control parameter from the vibration limit is more appropriate for determining the tetraquark masses, particularly for the open bottom families. This finding emphasizes the importance of choosing the correct control parameter for accurate mass spectroscopy.

Furthermore, we observed that in open bottom tetraquark states, the dominant contribution comes from the pairing of the third quarks, indicating that the phase parameters for these quarks play a crucial role in computing tetraquark masses at high energies around 5-6 GeV. In contrast, for open charm tetraquarks, there is a competition between the first and second quarks. These observations suggest that the pairing of quarks and the interplay of their phase parameters can significantly impact the mass spectroscopy of multiquark systems.

Our findings provide insights into the properties of fully-heavy tetraquarks and highlight the importance of considering the multiquark dynamics for a more comprehensive understanding of hadron spectroscopy. Further studies on other types of multiquarks and the inclusion of other degrees of freedom, such as pentaquarks and hexaquarks, may lead to a better understanding of the underlying physics and shed light on the nature of hadronic matter.

The comparison between our results obtained from the pairing model and the predictions of previous theoretical calculations are presented in Tables~\ref{tab:cm1} and~\ref{tab:cm2}. Our findings show that the pairing model provides reasonable agreement with the other works, implying that it has the potential to play a crucial role in predicting fully-heavy tetraquark mesons. However, future work could aim to further improve the understanding of multiquark dynamics by including the large-$N$ limit of the pure pairing Hamiltonian.


\section{Summary}
\label{sec:summary}

The aim of this study was to investigate the mass spectra of tetraquarks in the transition region between vibration and rotation using the algebraic framework. To achieve this, a solvable extended transitional Hamiltonian based on $SU(1,1)$ algebra is proposed, which can describe both partial high energy states and quantum phase transition. The extracted mass spectra of various tetraquarks were in agreement with previous research and other theoretical approaches. However, it is important to consider other degrees of freedom, such as penta or hexa quarks, in future studies. Furthermore, the solvable technique introduced in this work could potentially be applied to diagonalize more complex multiquark systems.
This approach is currently being applied to investigate other types of multiquarks in the following manuscript.


\begin{acknowledgments}
This work was supported in part by the National Natural Science Foundation of China (No.12250410254) and Polish National Science Centre (NCN) under Contract No 2018/31/B/ST2/02220 and . The Ministerio Espa\~nol de Ciencia e Innovaci\'on under grant no. PID2019-107844GB-C22; and Junta de Andaluc\'ia, contract nos. P18-FR-5057, Operativo FEDER Andaluc\'ia 2014-2020 UHU-1264517, and PAIDI FQM-370.
\end{acknowledgments}

\clearpage


\begin{thebibliography}{99}

\bibitem{lhcb1} R. Aaij et al. [LHCb], [arXiv:2009.00025 [hep-ex]].
\bibitem{lhcb2} R. Aaij et al. [LHCb], [arXiv:2009.00026 [hep-ex]].
\bibitem{pit} Pitaevskii, Lev, and Sandro Stringari. Bose-Einstein condensation and superfluidity. Vol. 164. Oxford University Press, (2016).
\bibitem{fra} Frauendorf, Stefan, and Augusto O. Macchiavelli. Progress in Particle and Nuclear Physics 78: 24-90  (2014).
\bibitem{leg}A. J. Leggett, Reviews of Modern Physics 73, 307 (2001).
\bibitem{arim}A. Arima and F. Iachello, Annals of Physics 281, 2 (2000).
\bibitem{oss91}F. Iachello, S. Oss, and R. Lemus, Journal of Molecular Spectroscopy 149, 132 (1991).
\bibitem{oss93}F. Iachello, S. Oss, and L. Viola, Molecular Physics 78, 561 (1993).
\bibitem{bar} Barabanov, M. Yu, et al.  Progress in Particle and Nuclear Physics 116: 103835 (2021).
\bibitem{kru} Krusche, B., and S. Schadmand.  Progress in Particle and Nuclear Physics 51.2: 399-485 (2003).
\bibitem{b2}R. Bijker, F. Iachello, and A. Leviatan, Annals of Physics 284, 89 (2000).
\bibitem{b1}R. Bijker, F. Iachello, and A. Leviatan, Annals of Physics 236, 69 (1994).
\bibitem{f1}F. Iachello, Nuclear Physics A 497, 23 (1989).
\bibitem{f2}F. Iachello, N. C. Mukhopadhyay, and L. Zhang, Phys. Rev. D 44, 898 (1991).
\bibitem{arima75} Arima, A., and F. Iachello.  Phys. Rev. Lett 35.16: 1069 (1975).
\bibitem{casten} Casten, Richard F., and David D. Warner. Reviews of Modern Physics 60.2: 389  (1988).
\bibitem{pan2002}F. Pan, X. Zhang, and J. Draayer, Journal of Physics A 35, 7173 (2002).
\bibitem{hu} Ui, Haruo. Annals of Physics 49.1 : 69-92 (1968).
\bibitem{pan2006}F. Pan, Y. Zhang, and J. Draayer, Eur. Phys. J. A 28, 313 (2006).
\bibitem{cap8} M. Caprio and F. Iachello, Nuclear physics A 781, 26 (2007).
\bibitem{27} M. Caprio, J. Skrabacz, F. Iachello,  Journal of Physics A: Mathematical and Theoretical, 44, 075303 (2011).
\bibitem{me16} A. Jalili. Majarshin, M. A. Jafarizadeh, and N. Fouladi, Eur. Phys. J. Plus 131, 1 (2016).
\bibitem{ajm17} A. J. Majarshin and M. Jafarizadeh, Nuclear Physics A  968: 287-325.(2017).
\bibitem{epja} A. Jalili Majarshin, Eur. Phys. J. A 54, 11 (2018).
\bibitem{ajm2021} A. J. Majarshin, Y.-A. Luo, F. Pan, and H. T. Fortune, Physical Review C 104, 014321 (2021).
\bibitem{ajmpt} A. J. Majarshin, Y.-A. Luo, F. Pan, H. Fortune, and J. P. Draayer, Physical Review C 103, 024317 (2021).
\bibitem{Karliner:2016zzc}M. Karliner, S. Nussinov, and J. L. Rosner, Phys. Rev. D 95, 034011 (2017).
\bibitem{Aaij:2018zrb}R. Aaij et al., Journal of High Energy Physics 7 (2018): 20.
\bibitem{dur}S. Durgut and C. Collaboration, in APS April Meeting Abstracts2018), p. U09. 006.
\bibitem{1804391}L. H. collaboration, Science Bulletin 65, 1983 (2020).
\bibitem{Berezhnoy:2011xn}A. Berezhnoy, A. Luchinsky, and A. Novoselov, arXiv preprint arXiv:1111.1867  (2011).
\bibitem{Wu:2016vtq}J. Wu, Y.-R. Liu, K. Chen, X. Liu, and S.-L. Zhu, Phys. Rev. D 97, 094015 (2018).
\bibitem{Chen:2016jxd}W. Chen, H.-X. Chen, X. Liu, T. G. Steele, and S.-L. Zhu, Phys. Lett. B 773, 247 (2017).
\bibitem{Wang:2017jtz}Z.-G. Wang, The European Physical Journal C 77, 432 (2017).
\bibitem{Wang:2018poa}Z.-G. Wang and Z.-Y. Di, arXiv preprint arXiv:1807.08520  (2018).
\bibitem{Reinders:1984sr}L. J. Reinders, H. Rubinstein, and S. Yazaki, Physics Reports 127, 1 (1985).
\bibitem{Heller:1985cb}L. Heller and J. A. Tjon, Physical Review D 32, 755 (1985).
\bibitem{Anwar:2017toa}M. N. Anwar, J. Ferretti, F.-K. Guo, E. Santopinto, and B.-S. Zou, Eur. Phys. J. C 78, 647 (2018).
\bibitem{Esposito:2018cwh}A. Esposito and A. D. Polosa, The European Physical Journal C 78, 782 (2018).
\bibitem{Ader:1981db}J. P. Ader, J. M. Richard, and P. Taxil, Physical Review D 25, 2370 (1982).
\bibitem{Zouzou:1986qh}S. Zouzou, B. Silvestre-Brac, C. Gignoux, and J. M. Richard, Zeitschrift für Physik C Particles and Fields 30, 457 (1986).
\bibitem{Lloyd:2003yc} R. J. Lloyd and J. P. Vary, Phys. Rev. D 70, 014009 (2004).
\bibitem{Barnea:2006sd}N. Barnea, J. Vijande, and A. Valcarce, Physical Review D 73, 054004 (2006).
\bibitem{Richard:2018yrm}J.-M. Richard, A. Valcarce, and J. Vijande, Physical Review C 97, 035211 (2018).
\bibitem{Richard:2017vry}J.-M. Richard, A. Valcarce, and J. Vijande, Physical Review D 95, 054019 (2017).
\bibitem{Vijande:2009kj}J. Vijande, A. Valcarce, and N. Barnea, Physical Review D 79, 074010 (2009).
\bibitem{Debastiani:2017msn}V. R. Debastiani and F. Navarra, Chinese Physics C 43, 013105 (2019).
\bibitem{Liu:2019zuc}M.-S. Liu, Q.-F. Lü, X.-H. Zhong, and Q. Zhao, Physical Review D 100, 016006 (2019).
\bibitem{Chen:2019dvd}X. Chen, The European Physical Journal A 55, 106 (2019).
\bibitem{Chen:2019vrj}X. Chen, Physical Review D 100, 094009 (2019).
\bibitem{Chen:2020lgj}X. Chen, arXiv preprint arXiv:2001.06755  (2020).
\bibitem{Wang:2019rdo}G.-J. Wang, L. Meng, and S.-L. Zhu, Physical Review D 100, 096013 (2019).
\bibitem{Yang:2020rih}G. Yang, J. Ping, L. He, and Q. Wang, arXiv preprint arXiv:2006.13756  (2020).
\bibitem{Bedolla:2019zwg}M. A. Bedolla, J. Ferretti, C. D. Roberts, and E. Santopinto, The European Physical Journal C 80, 1004 (2020).
\bibitem{Hughes:2017xie}C. Hughes, E. Eichten, and C. T. H. Davies, Physical Review D 97, 054505 (2018).
\bibitem{Gaudin76} M. Gaudin, Journal de Physique 37.10 (1976): 1087-1098.
\bibitem{p1}J. Wu, Y.-R. Liu, K. Chen, X. Liu, and S.-L. Zhu, Phys. Rev. D 97, 094015 (2018).
\bibitem{p2}W. Chen, H.-X. Chen, X. Liu, T. G. Steele, and S.-L. Zhu, Phys. Lett. B 773, 247 (2017).
\bibitem{p4}P. Lundhammar and T. Ohlsson, Phys. Rev. D 102, 054018 (2020).
\bibitem{p5}C. Deng, H. Chen, and J. Ping, Phys. Rev. D 103, 014001 (2021).
\bibitem{zha}J. Zhao, S. Shi, and P. Zhuang, Phys. Rev. D 102, 114001 (2020).
\bibitem{yang}G. Yang, J. Ping, and J. Segovia, Symmetry 12, 1869 (2020).
\bibitem{FullBeauty2019}Y. Bai, S. Lu, and J. Osborne, Phys. Lett. B 798, 134930 (2019).
\bibitem{FullHeavy2019sec}M. A. Bedolla, J. Ferretti, C. D. Roberts, and E. Santopinto, Eur. Phys. J. C 80, 1004 (2020).
\bibitem{FullHeavy2018}M. N. Anwar, J. Ferretti, F.-K. Guo, E. Santopinto, and B.-S. Zou, Eur. Phys. J. C 78, 647 (2018).
\bibitem{D12}A. V. Berezhnoy, A. V. Luchinsky, and A. A. Novoselov, Phys. Rev. D 86, 034004 (2012).
\bibitem{blln}A. Berezhnoy, A. Likhoded, A. Luchinsky, and A. Novoselov, Phys. At. Nucl.75, 1006 (2012).
\bibitem{FullHeavy2017}M. Karliner, S. Nussinov, and J. L. Rosner, Phys. Rev. D 95, 034011 (2017).
\bibitem{Karliner:2020dta}M. Karliner and J. L. Rosner, Phys. Rev. D 102, 114039 (2020).
\bibitem{SumR2}Z.-G. Wang, Eur. Phys. J. C 77, 432 (2017).
\bibitem{E18}Z.-G. Wang and Z.-Y. Di, Acta Phys. Pol. B 50, 1335 (2019).
\bibitem{Chen}X. Chen,  Eur. Phys. J. A 55, 106 (2019).
\bibitem{Jin:2020jfc}X. Jin, Y. Xue, H. Huang, and J. Ping, Eur. Phys. J. C 80, 1083 (2020).
\bibitem{WLZ}G.-J. Wang, L. Meng, and S.-L. Zhu,  Phys. Rev. D 100, 096013 (2019).
\bibitem{FullHeavy2019}M.-S. Liu, Q.-F. Lu, X.-H. Zhong, and Q. Zhao, Phys. Rev. D 100, 016006 (2019).
\bibitem{liu:2020eha}M.-S. Liu, F.-X. Liu, X.-H. Zhong, and Q. Zhao, Full-heavy tetraquark states and their evidences in the LHCb di-$J/\psi$ spectrum,arXiv:2006.11952.
\bibitem{Lu:2020cns}Q.-F. L\"u, D.-Y. Chen, and Y.-B, Eur. Phys. J. C 80, 871 (2020).
\bibitem{FullCharm2017}V. R. Debastiani and F. S. Navarra, Spectroscopy of the all charm tetraquark, Proc. Sci., Hadron 2017 (2018) 238.
\bibitem{E19}V. R. Debastiani and F. S. Navarra, Chin. Phys. C 43, 013105 (2019).
\bibitem{102} R. J. Lloyd and J. P. Vary, Phys. Rev. D 70, 014009 (2004).
\bibitem{103}Z. Zhao, K. Xu, A. Kaewsnod, X. Liu, A. Limphirat, and Y. Yan, arXiv preprint arXiv:2012.15554  (2020).
\bibitem{tetrababc}H. Mutuk, Eur. Phys. J. C 81, 367 (2021).
\bibitem{tetrac}X. Chen, arXiv preprint arXiv:2001.06755  (2020).
\bibitem{25}X.-Z. Weng, X.-L. Chen, W.-Z. Deng, and S.-L. Zhu, Phys. Rev. D 103, 034001 (2021).
\bibitem{26}R. N. Faustov, V. O. Galkin, and E. M. Savchenko, Phys. Rev. D 102, 114030 (2020).
\bibitem{Chen2}X. Chen,  Phys. Rev. D 100, 094009 (2019).
\bibitem{lu} L\"u, Qi-Fang, and Yu-Bing Dong. Physical Review D 94.9: 094041  (2016).
\bibitem{ebert} Ebert, D., R. N. Faustov, and V. O. Galkin.  Physics Letters B 696.3: 241-245 (2011).
\bibitem{pg} Particle Data Group, J. Phys. G 37 (2010) 075021.












  \end{thebibliography}
\end{document}